\documentclass[sigconf]{acmart}
\copyrightyear{2024}
\acmYear{2024}
\setcopyright{rightsretained}
\acmConference[ICSE '24]{2024 IEEE/ACM 46th International Conference on Software Engineering}{April 14--20, 2024}{Lisbon, Portugal}
\acmBooktitle{2024 IEEE/ACM 46th International Conference on Software Engineering (ICSE '24), April 14--20, 2024, Lisbon, Portugal}
\acmDOI{10.1145/3597503.3639076}
\acmISBN{979-8-4007-0217-4/24/04}
\usepackage{tikz}
\usetikzlibrary{arrows.meta}
\usepackage{amsmath}
\usepackage{xspace}
\usepackage{balance}
\usepackage{color, xcolor}
\usepackage{threeparttable}
\usepackage{listings}
\usepackage{foreign}
\usepackage[normalem]{ulem}

\usepackage{tcolorbox}
\definecolor{mycolor}{rgb}{0.122, 0.435, 0.698}
\definecolor{gray1}{gray}{0.3}
\definecolor{darkgreen}{rgb}{0.0, 0.5, 0.0}
\definecolor{darkred}{rgb}{0.82, 0.1, 0.26}
\definecolor{shallowgreen}{RGB}{196, 214, 160}
\definecolor{shallowred}{RGB}{217, 149, 143}
\newcommand{\result}[1]{%
\begin{tcolorbox}[colframe=mycolor,boxrule=0.5pt,arc=4pt,
      left=6pt,right=6pt,top=6pt,bottom=6pt,boxsep=0pt,width=\columnwidth]%
      {#1}
\end{tcolorbox}%
}

\newcounter{tmlistings}
\newcommand\makenode[2]{%
  \tikz[baseline=0pt, remember picture] {\node (listings-\the\value{tmlistings}) {#2}; }%
  \stepcounter{tmlistings}%
}

\newcommand{\sqlancer}{\textsl{SQLancer}\xspace}
\newcommand{\method}{\textsl{CERT}\xspace} 
\newcommand{\tool}{\textsl{SQLancer+CERT}\xspace}
\newcommand{\amoeba}{\textsl{AMOEBA}\xspace}
\newcommand{\apollo}{\textsl{APOLLO}\xspace}
\newcommand{\property}{\emph{cardinality restriction monotonicity}\xspace}
\newcommand{\todo}[1]{}
\renewcommand{\todo}[1]{{\color{red} TODO: {#1}}}
\newcommand{\add}[1]{}
\renewcommand{\add}[1]{{\color{red} {#1}}}
\newcommand{\del}[1]{}
\renewcommand{\del}[1]{{\color{blue} \sout{#1}}}

\newcommand{\numreported}{14\xspace}
\newcommand{\numbugs}{13\xspace}
\newcommand{\numconfirmedbugs}{9\xspace}
\newcommand{\numunknownbugs}{12\xspace}
\newcommand{\numfixedbugs}{2\xspace}
\newcommand{\confirmedrate}{85\%\xspace}
\newcommand{\amoebaconfirmed}{6\xspace}
\newcommand{\amoebaconfirmedrate}{24\%\xspace}
\newcommand{\speedfactor}{386$\times$\xspace}

\newcommand{\violatedpercentage}{0.2\%\xspace}

\newcommand{\etal}{\textit{et al}.\xspace}

\newcommand{\bugsymbol}[0]{\includegraphics[width=0.3cm]{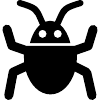}}
\newcommand{\oksymbol}[0]{\includegraphics[width=0.3cm]{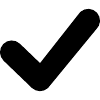}}

\newcommand{\Autoref}[1]{%
  \begingroup%
  \def\chapterautorefname{Chapter}%
  \def\sectionautorefname{Section}%
  \def\subsectionautorefname{Subsection}%
  \autoref{#1}%
  \endgroup%
}



\begin{document}
\title{CERT: Finding Performance Issues in Database Systems Through the Lens of Cardinality Estimation}

\author{Jinsheng Ba}
\orcid{0000-0003-4008-9225}
\affiliation{%
  \institution{National University of Singapore}
  \country{Singapore}
}
\email{bajinsheng@u.nus.edu}

\author{Manuel Rigger}
\orcid{0000-0001-8303-2099}
\affiliation{%
  \institution{National University of Singapore}
  \country{Singapore}
}
\email{rigger@nus.edu.sg}

\begin{abstract}
Database Management Systems (DBMSs) process a given query by creating a query plan, which is subsequently executed, to compute the query's result. Deriving an efficient query plan is challenging, and both academia and industry have invested decades into researching query optimization.
Despite this, DBMSs are prone to performance issues, where a DBMS produces an unexpectedly inefficient query plan that might lead to the slow execution of a query.
Finding such issues is a longstanding problem and inherently difficult, because no ground truth information on an expected execution time exists.
In this work, we propose \emph{Cardinality Estimation Restriction Testing} (\method), a novel technique that finds performance issues through the lens of cardinality estimation.
Given a query on a database, \method derives a more restrictive query (\emph{e.g.}, by replacing a \lstinline{LEFT JOIN} with an \lstinline{INNER JOIN}), whose estimated number of rows should not exceed the estimated number of rows for the original query.
\method tests cardinality estimation specifically, because it was shown to be the most important part for query optimization; thus, we expect that finding and fixing cardinality-estimation issues might result in the highest performance gains. In addition, we found that other kinds of query optimization issues can be exposed by unexpected estimated cardinalities, which can also be found by \method.
\method is a black-box technique that does not require access to the source code; DBMSs expose query plans via the \lstinline{EXPLAIN} statement.
\method eschews executing queries, which is costly and prone to performance fluctuations.
We evaluated \method on three widely used and mature DBMSs, MySQL, TiDB, and CockroachDB. 
\method found \numbugs unique issues, of which \numfixedbugs issues were fixed and \numconfirmedbugs confirmed by the developers.
We expect that this new angle on finding performance bugs will help DBMS developers in improving DMBSs' performance.
\end{abstract}

\maketitle

\section{Introduction}

Database Management Systems (DBMSs) are fundamental software systems that allow users to retrieve, update, and manage data~\cite{laney20013d, oracle, stonebraker2013intel}.
Most DBMSs support the \emph{Structured Query Langage} (SQL), which allows users to specify queries in a declarative way.
Subsequently, the DBMSs translate the query into a concrete execution plan.
To balance the trade-off between spending little time on optimization, which is performed at run time, and finding an efficient execution plan, researchers and practitioners have invested decades of effort into query optimization, covering directions such as search space exploration for join ordering~\cite{neumann2009query, fender2012effective, fender2013counter}, index data structures~\cite{graefe2011modern}, execution time prediction~\cite{akdere2012learning, wu2013predicting}, or parallel execution on multi-core CPUs~\cite{giceva2014deployment} and GPUs~\cite{paul2016gpl}.

Finding performance issues in DBMSs---also referred to as optimization opportunities or performance bugs---is challenging.
Given a query $Q$ and a database $D$, we want to determine whether executing $Q$ on $D$ results in unexpectedly suboptimal performance. 
In general, no ground truth is available that specifies whether $Q$ executes within a reasonable time.
To exacerbate this issue, DBMSs use various heuristics and cost models during optimizations, or make trade-offs in optimizing specific kinds of queries over others.
Second, the execution time of $Q$ might be significant if $D$ is large, making it time-consuming to measure $Q$'s actual performance.
Given that the execution time depends on various factors of the execution environment~\cite{mytkowicz2009producing} (\emph{e.g.}, the state of caches), it might even be necessary to execute $Q$ multiple times to obtain a reasonably reliable measure of its execution time.
Cloud environments are in particular prone to noise~\cite{laaber2019software}; a report on testing SAP HANA~\cite{bach2022testing} has recently stressed that performance testing for cloud offerings of DBMSs---such as SAP HANA Cloud, which runs in Kubernetes pods---is one of the main challenges in testing DBMSs due to inherently noisy environments.

Benchmark suites such as TPC-DS~\cite{tpcds} or TPC-H~\cite{tpch} are widely used in practice to monitor DBMSs' performance over versions through predetermined performance baselines, which could be specified~\cite{rehmann2016performance, yagoub2008oracle, yan2018snowtrail}. However, deriving an appropriate baseline is challenging and might result in false alarms. Automated testing techniques have been proposed to find performance issues without the need of curating a benchmark suite.
\apollo~\cite{jung2019apollo} generates databases and queries automatically and validates whether executing the query on different versions of the DBMS results in significantly different execution times. However, \apollo can only find regression bugs.
\amoeba~\cite{liu2022automatic} finds performance issues by examining discrepancies in the execution time of a pair of semantically equivalent queries. However, semantically equivalent queries do not necessarily exhibit a similar performance as the issues found by \amoeba have a high false positive rate---only 6 of 39 issues were confirmed by the developers, 5 of which were fixed~\cite{liu2022automatic}.
For the above methods, queries need to be executed on sufficiently large databases to detect significant performance discrepancies.

In this work, we propose \emph{Cardinality Estimation Restriction Testing} (\method), a general technique that finds performance issues by testing the DBMSs' cardinality estimation.
\emph{Cardinality estimation} is the process in which \emph{cardinality estimator} computes \emph{estimated cardinalities}, the estimated numbers of rows that will be returned.
Since estimated cardinalities are approximate, it is infeasible to check for a specific number.
Rather, the core idea of our approach is that making a given query more restrictive should cause the cardinality estimator to estimate that the more restrictive query should fetch at most as many rows as the original query. More formally, given a query $Q$ and a database $D$, $Card(Q, D)$ denotes the actual cardinality, that is, the exact number of records to be fetched by $Q$ on $D$. If we derive a more restrictive query $Q'$ from $Q$, $Card(Q', D) \leq Card(Q, D)$ always holds. $EstCard(Q, D)$ denotes the estimated cardinality for $Q$, and we expect $EstCard(Q', D) \leq EstCard(Q, D)$ to also hold for any DBMS. We refer to this property as \property. Any violation of this property indicates a potential performance issue.

\method addresses the aforementioned challenges. Cardinality estimation accuracy was shown to be the single most important component for deriving an efficient execution plan~\cite{leis2015good}. Therefore, we believe that pinpointing issues in cardinality estimation would help developers focus on the most relevant issues, addressing which might result in significant performance gains. Additionally, this idea is applicable to finding a broader range of performance issues. For example, we found that other kinds of query optimization issues can be exposed by unexpected estimated cardinalities, as shown in \autoref{lst:prebug}. Estimated cardinalities can be readily obtained by DBMSs without executing $Q$; DBMSs typically provide a SQL \lstinline{EXPLAIN} statement that provides this information as part of a query plan, allowing our technique to achieve high throughput. Furthermore, since our method does not measure run-time performance, it can be used in noisy environments, and minimal test cases that demonstrate the performance issue can be automatically obtained~\cite{zeller2009programs}. Finally, \method is a black-box technique, because applying it requires no access to the source code or binary.

\Autoref{lst:motivation} shows a running example demonstrating \method. We randomly generate SQL statements as shown in lines 1--4 to create a database state and ensure that each table's data statistics are up to date in lines 5--6. Then, we randomly generate a query with a \lstinline{LEFT JOIN} and derive a more restrictive query by replacing the \lstinline{LEFT JOIN} with an \lstinline{INNER JOIN} as shown in lines 8--9. The second query is more restrictive than the first query as \lstinline{INNER JOIN} should always fetch no more rows than \lstinline{LEFT JOIN}. 
We examined the estimated cardinalities in their query plans, obtained by using an \lstinline{EXPLAIN} statement.
Despite having made the query more restrictive, the cardinality estimator estimated that the first query fetches 20 rows, while the second one fetches 60 rows, which is unexpected. The root cause was an incorrect double-counting when estimating the selectivity of \lstinline{OR} expression in the \lstinline{ON} condition of the \lstinline{INNER JOIN}. The estimated cardinality of the first query with \lstinline{LEFT JOIN} should be no less than 60; after the developers fixed this issue, the estimated cardinality was changed to 60. This fix improved the performance of the query \lstinline{SELECT * FROM t0 LEFT OUTER JOIN t1 ON t0.c0<1 OR t0.c0>1 FULL JOIN t2 ON t0.c0=t2.c0} by 20\% as shown in \Autoref{lst:case3}. The improvement was due to a more accurate estimated cardinality, which enabled a better selection of the join order. Note that we avoided executing the query; \method only examines query plans.

\begin{figure}
\begin{lstlisting}[caption={This running example demonstrates a performance issue found by \method in CockroachDB. It is due to an incorrect double-counting when estimating the selectivity of OR expressions in join ON conditions.},captionpos=t, label=lst:motivation, escapeinside=&&]
CREATE TABLE t0 (c0 INT);
CREATE TABLE t1 (c0 INT);
INSERT INTO t0 VALUES (1), (2), (3), (4), (5), (6), (7), (8), (9), (10), (11), (12), (13);
INSERT INTO t1 VALUES (21),(22),(23),(24),(25);  
ANALYZE t0; 
ANALYZE t1;

EXPLAIN SELECT * FROM t0 &\underline{\textbf{LEFT JOIN}}& t1 ON t0.c0<1 OR t0.c0>1; -- estimated rows: 20 &\bugsymbol&
EXPLAIN SELECT * FROM t0 &\underline{\textbf{INNER JOIN}}& t1 ON t0.c0<1 OR t0.c0>1; -- estimated rows: 60 &\oksymbol&
-------------------------------------------------
&$\bullet$& cross join(left outer) &$\bullet$& cross join
| estimated row:20       | estimated row:60
| pred:(c0<1)OR(c0>1)    |-&$\bullet$& filter
|-&$\bullet$& scan                 |  | estimated row:12
|    estimated row:13    |  | filter:(c0<1)OR(c0>1)
|    table: t0@t0_pkey   |  |-&$\bullet$& scan
|-&$\bullet$& scan                 |       estimated row:13
     estimated row:5     |       table: t0@t0_pkey
     table: t1@t1_pkey   |-&$\bullet$& scan
                              estimated row:5
                              table: t1@t1_pkey

\end{lstlisting}
\end{figure}

We implemented \method in \sqlancer, a popular DBMS testing tool, and evaluated it on three widely used and mature DBMSs, MySQL, TiDB, and CockroachDB.
While MySQL is one of the most popular open-source DBMSs, TiDB and CockroachDB are developed by companies.
We reported \numreported performance issues to the developers, who confirmed that \numbugs of them were unique and \numunknownbugs were unknown. Of these unique issues, \numfixedbugs issues were fixed, \numconfirmedbugs other issues were confirmed, and 2 issues required further investigation.
Similar to existing work, \method might report false alarms, since implementations might not strictly adhere to the \property. However, in practice, none of the issues that we reported were considered false alarms.
Our evaluation demonstrates the high throughput achieved by eschewing executing queries; our implementation can validate \speedfactor more queries than \amoeba in the same time period.
We believe that these results demonstrate that \method might become a standard technique in DBMS developers' toolbox, due to its efficiency and effectiveness, and hope that it will inspire future work on finding performance issues in DBMSs.

Overall, we make the following contributions:

\begin{itemize}
    \item We present a motivational study to investigate the causes of previous performance issues.
    \item We propose a novel technique, \method, to test cardinality estimation for finding performance issues in query optimization without measuring execution time. We show a concrete realization of the technique by proposing 12 query-restriction rules.
    \item We implemented \method in \sqlancer and evaluated it on multiple aspects. \method found \numbugs unique issues of cardinality estimation in widely-used DBMSs, and 11 issues were confirmed or fixed. The source code of \method is publicly available, and has been integrated into \sqlancer.\footnote{\url{https://github.com/sqlancer/sqlancer/issues/822}}
\end{itemize}

\section{Background}

\paragraph{Structured Query Language.}
\sloppy{}
Structured Query Language (SQL)~\cite{chamberlin1974sequel} is a declarative programming language that expresses only the logic of a computation without specifying its specific execution. SQL is widely supported by DBMSs; for example, according to a popular ranking,\footnote{\url{https://db-engines.com/en/ranking} as of March 2023.} the 10 most popular DBMSs support it. \Autoref{lst:select} shows the EBNF representation~\cite{feynman2016ebnf}, a metasyntax notation to express context-free grammars, of a SQL query, whose features we considered in this work. A query starts with the \lstinline{SELECT} keyword. It can optionally be succeeded by a \lstinline{DISTINCT} clause that specifies that only unique records should be returned. A \lstinline{JOIN} clause joins two tables or views; various joins exist that differ on whether and what rows should be joined when the join predicate evaluates to false. A query can contain a single \lstinline{WHERE} clause; only rows for which its predicate evaluates to true are included in the result set. Similar to \lstinline{DISTINCT}, the \lstinline{GROUP BY} clause groups rows that have the same values into a single row. It can be followed by a \lstinline{HAVING} clause that excludes records after grouping them. The \lstinline{LIMIT} clause is used to restrict the number of records that are fetched.
More advanced features, such as window functions, common table expressions (CTEs), and subqueries can be used. While we did not consider them in this work, we believe that our proposed approach could be extended to support them.

\begin{figure}
\begin{lstlisting}[caption={The EBNF representation of a query.},captionpos=t, label=lst:select, escapeinside=&&]
SELECT [DISTINCT] 
  select_expression (, select_expression)*
  FROM table_reference (INNER | LEFT | RIGHT | FULL | CROSS JOIN table_reference)*
  (WHERE predicate)+
  (GROUP BY predicate
  (HAVING predicate)+)+
  (LIMIT row_count)+ ;
\end{lstlisting}
\end{figure}

\paragraph{Query optimization.}  DBMSs include query optimizers that, after parsing the SQL query, determine an efficient \emph{query plan}, which specifies how a SQL query is executed. Determining an efficient query plan is challenging, since many factors might influence the plan's performance. The most commonly used models are cost-based~\cite{chaudhuri1998overview}---the query plan with the lowest projected performance cost is chosen. Cardinality estimation was found to be the most important factor that affects the quality of query optimization~\cite{leis2015good}. Cardinality estimators typically obtain data statistics of the tables to be queried by sampling~\cite{heimel2015self}, through histograms~\cite{selinger1979access}, or machine learning algorithms~\cite{dutt2019selectivity, wu2021unified, kipf2018learned}. Then, they enumerate all sub-plan queries, which are queries that process only a subset of tables in a query, and estimate how many rows they fetch. For example, for a query $A \Join B \Join C$ ($\Join$ denotes a join), cardinality estimators could estimate the cardinalities of $A, B, C$ respectively, and then estimate the cardinalities of $A \Join B$, $B \Join C$, $A \Join C$, and $A \Join B \Join C$. Lastly, the estimated cardinalities of these sub-plan queries help to decide the join order---whether $A \Join B$ or $B \Join C$ should be executed first.

\paragraph{Query plan.} A query plan is a tree of operations that describes how a specific DBMS executes a SQL statement. A query plan can be obtained by executing a query with the prefix \lstinline{EXPLAIN}. Query plans typically include estimated cardinalities for operations that affect the number of subsequent rows. \Autoref{lst:motivation} shows the two query plans for the two queries of the running example in lines 11--21. The first query uses a \lstinline{LEFT JOIN}, and its query plan includes three operations: \lstinline{cross join}, \lstinline{scan}, and \lstinline{scan}. The second query uses an \lstinline{INNER JOIN}, and its query plan includes four operations: \lstinline{cross join}, \lstinline{filter}, \lstinline{scan}, and \lstinline{scan}. The structural difference between both query plans is the location of operation \lstinline{filter}, as the predicate \lstinline{(c0<1)OR(c0>1)} in the \lstinline{ON} clause of both queries is part of the \lstinline{cross join} of the first query plan, but is a separate operation \lstinline{filter} in the second query plan. For the first query plan, the estimated cardinality 20 of the root node \lstinline{cross join (left outer)} is determined based on the predicate as well as the two estimates for the number of records in tables \lstinline{t0} and \lstinline{t1}, which are 13 and 5. For the second query plan, the estimated cardinality for \lstinline{cross join} is 60; here, the estimate is partly based on the \lstinline{filter} operation, which is estimated to return 12 rows. In this work, when we refer to the estimated cardinality of a query, we refer to the query's root node.
\section{Performance Issue Study}\label{sec:study}

As a motivating study, to investigate if performance issues are caused by incorrect cardinality estimation in practice, we examined previous performance issues related to query optimization. 

\paragraph{Subjects.} We studied the issues reported for MySQL, TiDB, and CockroachDB. MySQL is the most popular relational DBMS according to a survey in 2021.\footnote{\url{https://insights.stackoverflow.com/survey/2021\#most-popular-technologies-database}} TiDB and CockroachDB are popular enterprise-class DBMSs, and their open versions on GitHub are highly popular as they have been starred more than 33k and 26k times. They are widely used and have thus been studied in other DBMS testing works~\cite{liang:sqlright, Rigger2020PQS, Rigger2020TLP}.

\paragraph{Methodology.} We searched for performance issues using the keywords "\emph{slow}" or "\emph{suboptimal}" in the above-stated DBMSs, aiming to obtain issues that relate to either slow execution or suboptimal query plans. For MySQL, we chose issues whose status was \emph{closed}, the severity was \emph{(S5) performance}, and the type was \emph{MySQL Server: Optimizer} in the bug tracker.\footnote{\url{https://bugs.mysql.com/}} Considering MySQL was first released in 1995 and some issues are too old to be reproduced, we investigated the issues in version 5.5 or later. For TiDB, we searched its repository\footnote{\url{https://github.com/pingcap/tidb/issues}} by the filter \emph{is:issue is:closed linked:pr label:type/bug slow in:title}. For CockroachDB, we searched its repository\footnote{\url{https://github.com/cockroachdb/cockroach/issues}} by the filter \emph{is:issue is:closed linked:pr label:C-bug slow in:title}. Then, we manually analyzed and reproduced each issue to identify whether it is a performance issue related to query optimization and caused by cardinality estimation.
Specifically, if the estimated cardinality of the query in a report was changed by the fix, we deemed the performance issue to be caused by incorrect cardinality estimation.

\begin{table}[]
    \centering\small
    \caption{Previous performance issues.}
    \label{tab:study}
    \begin{tabular}{@{}lrr@{}}
    \toprule
    \textbf{DBMS} & \textbf{\#ID} & \textbf{Caused by Cardinality Estimation} \\
    \midrule
    MySQL 	     & 61631  & \checkmark  \\
    MySQL 	     & 56714  & \checkmark  \\
    MySQL 	     & 25130  & \text{\sffamily X}  \\
    CockroachDB	 & 93410  & \text{\sffamily X}  \\
    CockroachDB	 & 71790  & \text{\sffamily X}  \\
    TiDB	     & 9067   & \checkmark  \\
    \bottomrule
    \textbf{Sum} & 6  & 3
    \end{tabular}
\end{table}

\paragraph{Analysis.} \autoref{tab:study} shows the studied performance issues. Overall, we identified six issues in three DBMSs, and three of them were caused by incorrect cardinality estimation. We attribute the lower number of performance issues to the difficulty in identifying and resolving performance issues in query optimization. For issues \#61631 and \#56714, they produce inefficient query plans that have higher estimated cardinalities than the optimal query plans. Although both issues are not directly due to the faults in cardinality estimators, they are still observable through estimated cardinalities, so we deemed both were caused by cardinality estimation. Issue \#9067 was caused by the cardinality estimation due to an issue in calculating cardinality for correlated columns. The other three issues, which were not caused by cardinality estimation, were due to inefficient operations. For example, issue \#71790 was due to the inefficient implementation of \lstinline{MERGE JOIN}, which did not use the smaller table as the right child, and the estimated cardinality remained unchanged after fixing the implementation of the operation.

\begin{figure}
\begin{lstlisting}[caption={The performance issue \#56714 in MySQL.},captionpos=t, label=lst:prebug, escapeinside=&&]
CREATE TABLE test (a INT PRIMARY KEY AUTO_INCREMENT, b INT NOT NULL, INDEX (b)) engine=INNODB;
CREATE TABLE integers(i INT UNSIGNED NOT NULL);
INSERT INTO integers(i) VALUES (0), (1), (2), (3), (4), (5), (6), (7), (8), (9);
INSERT INTO test (b)
SELECT units.i MOD 2
FROM integers AS units
    CROSS JOIN integers AS tens
    CROSS JOIN integers AS hundreds
    CROSS JOIN integers AS thousands
    CROSS JOIN integers AS tenthousands
    CROSS JOIN integers AS hundredthousands;
    
EXPLAIN SELECT MAX(a) FROM test WHERE b=0; -- estimated rows: {500360} &\bugsymbol&, {1} &\oksymbol&
\end{lstlisting}
\end{figure}

\paragraph{Case study.} \autoref{lst:prebug} shows issue  \#56714 in MySQL as an illustrative example of a performance issue caused by cardinality estimation. According to the issue report, this performance issue incurs a slowdown of execution time from 0.01 seconds to 3.02 seconds. Column \lstinline{b} in table \lstinline{test} uses an index, but the query in line 13 does not correctly use the index, incurring a \lstinline{FULL TABLE SCAN}, which is inefficient. Although the root cause for this performance is in index selection, not in the cardinality estimator, the suboptimal index selection affects the estimated cardinality as the \lstinline{FULL TABLE SCAN} is expected to scan more rows than the \lstinline{INDEX SCAN}. 

\result{Performance issues can arise from inefficient operations, flawed cardinality estimators, and inefficient query plans. The latter two causes can be found by unexpected estimated cardinalities.}
\section{Approach}

\begin{figure*}
    \centering
    \includegraphics[width=\textwidth]{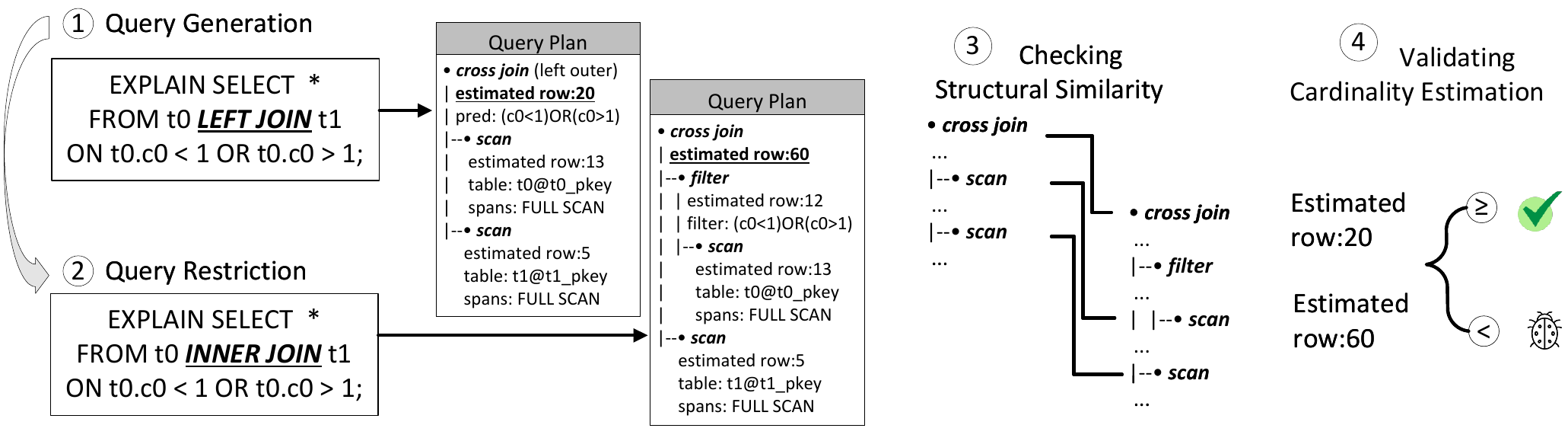}
    \caption{Overview of \method.}
    \label{fig:approach}
\end{figure*}

We propose \method, a novel technique for testing cardinality estimation. The core idea is that a given query should not have a lower estimated cardinality than a more restrictive query derived from it. We term this property \property and expect that DBMSs adhere to it in practice. \method is a simple black-box technique, making it widely applicable in practice.

\paragraph{Method overview.} \autoref{fig:approach} shows an overview of \method based on the running example from \autoref{lst:motivation}. Given a randomly generated query in step~\textcircled{1}, we derive another more restrictive query in step~\textcircled{2} and retrieve both queries' query plans. Then, if both query plans are structurally similar in step~\textcircled{3}, we validate the \property property in step~\textcircled{4}; we expect the less restrictive query from step \textcircled{1} to return at least as high estimated cardinality as the more restrictive query from step~\textcircled{2}. Any discrepancy is considered a performance issue. We perform the structural similarity checking in step~\textcircled{3} based on the observation that a more restrictive query can result in a significantly different query plan, whose estimated cardinalities are not comparable. Next, we give a detailed explanation of each step.

\subsection{Database and Query Generation} 
We require a database state and a query for testing. Both database state and query can be manually provided or generated. Common generation-based methods include mutation-based methods~\cite{zhong2020squirrel, liang:sqlright} and rule-based generation methods~\cite{sqlsmith, Rigger2020TLP, Rigger2020PQS, Rigger2020NoREC}. How to generate database states and queries is not a contribution of this paper, and our approach can be paired with any database state and query generation method.
For example, in \Autoref{lst:motivation}, we could randomly generate a database state in lines 1--4 and a query in line 8.

Before executing queries on the generated database state, we execute \lstinline{ANALYZE} statements on each table to guarantee that the data statistics are up to date. For \Autoref{lst:motivation}, these statements are executed in lines 5--6.

\subsection{Query Restriction}\label{subsec:queryrestriction}

Given a query, we derive a more restrictive query based on two insights. First, for the clauses that we considered, adding a clause to a query makes the query more restrictive except for the \lstinline{JOIN} clause.
Second, given an already existing clause, we can modify the clause or its predicate to obtain a more restrictive query.
We considered the SQL features shown in \Autoref{lst:select} and propose at least one rule for each feature, yielding the 12 rules shown in \Autoref{tab:rules}.
Since the \lstinline{JOIN} clause, which specifies two tables or views to be joined, is a major factor influencing the queries' run time~\cite{leis2015good}, 5 of the 12 rules relate to them.
Our rules are not exhaustive; we believe that practitioners could propose additional rules depending on their testing focus.

\paragraph{Rule overview.}
In \Autoref{tab:rules}, \method derives a \emph{Target} statement from a \emph{Source} statement by applying a restriction on the shown \emph{Clause}, as demonstrated through \emph{Example}. <Predicate> refers to a boolean expression and <Natural number> to a natural number. These examples are based on a database state with two tables \lstinline{t0} and \lstinline{t1}, both of which have only one column \lstinline{c0}. 
For each test to be generated, we randomly choose a SQL clause, of which one or more rules are randomly applied to restrict a query. 
In \Autoref{fig:approach}, we choose the \emph{JOIN} clause and apply only rule 1, which replaces a \lstinline{LEFT JOIN} with an \lstinline{INNER JOIN} in the \lstinline{JOIN} clause of a query.

\begin{figure}
    \centering
    \includegraphics[width=0.9\columnwidth]{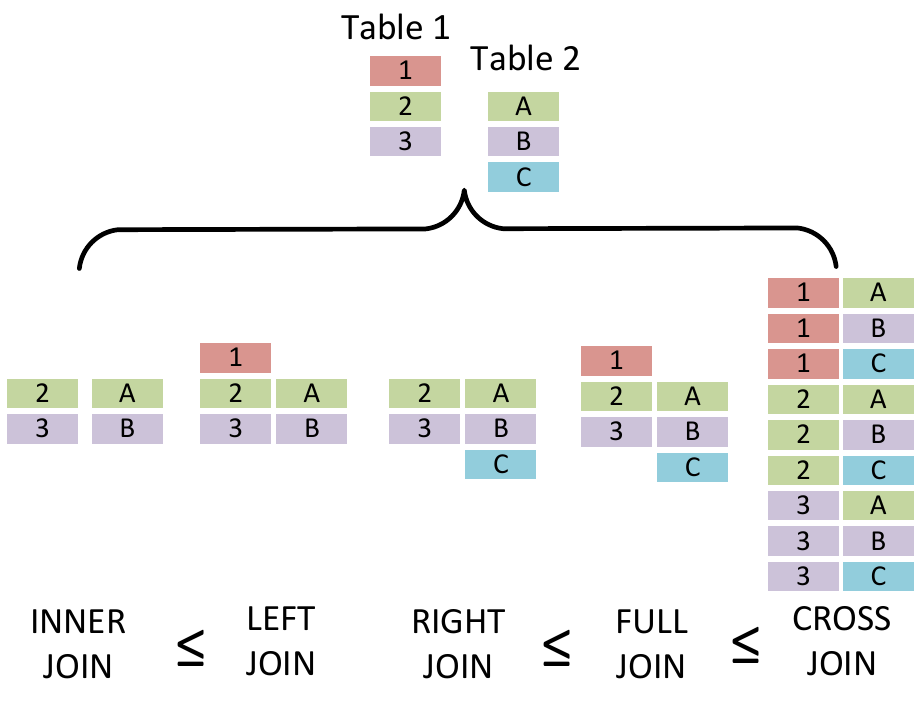}
    \caption{The inequality relationships of estimated cardinalities in the \lstinline{JOIN} clause with an example to join two tables.}
    \label{fig:join}
\end{figure}

\paragraph{JOIN clause.} Our key insight for testing the \lstinline{JOIN} clause is the partial inequality relationship in terms of cardinalities between different kinds of joins.
For a fixed join predicate, the following inequalities for the different joins' cardinalities hold:
\lstinline{INNER JOIN} $\leq$ \lstinline{LEFT JOIN/RIGHT JOIN} $\leq$ \lstinline{FULL JOIN} $\leq$ \lstinline{CROSS JOIN}. \Autoref{fig:join} illustrates this using a \lstinline{JOIN} diagram~\cite{explainjoin} on two tables, each of which has three rows, with the same color denoting the rows that can be matched in the \lstinline{JOIN} predicate. As determined by the SQL standard, \lstinline{INNER JOIN} fetches rows that have matching values in both tables; \lstinline{LEFT JOIN/RIGHT JOIN} fetch all rows from the left/right table and the matching rows from the respectively other table; \lstinline{FULL JOIN} fetches all rows from both tables; \lstinline{CROSS JOIN} fetches all possible combinations of all rows from both tables without an \lstinline{ON} clause. 
A corner case for rule 5 concerning the \lstinline{CROSS JOIN} is that this join may fetch fewer rows than \lstinline{FULL JOIN} if either of the tables is empty, in which case \lstinline{CROSS JOIN} fetches zero rows. To avoid potential false alarms, we ensure that each table contains at least one row.

\paragraph{WHERE clause.} 
For the \lstinline{WHERE} clause, our insight is that we can restrict the predicate that is used for filtering rows. 
If the query contains no \lstinline{WHERE} clause, we restrict the query by adding one with a random predicate.
If the \lstinline{WHERE} clause's predicate has an \lstinline{OR} operator, we restrict it by removing either of the \lstinline{OR}'s operands. Otherwise, we add an \lstinline{AND} operator with a randomly generated predicate.
Restricting predicates would also apply to testing \lstinline{JOIN} clauses; in this work, we aimed to introduce the general idea behind \method and illustrate it on a small set of promising rules. We believe that practitioners who adopt the approach will propose many additional rules.

\paragraph{Other SQL clauses.}
A query can be restricted by a \lstinline{DISTINCT} clause, which should fetch no more rows than the same query without such a clause, or by replacing its \lstinline{ALL} clause. Similarly, a query without \lstinline{GROUP BY} or \lstinline{HAVING} can be restricted by adding such clauses along with any predicate. A \lstinline{LIMIT} clause can be added, or a lower limit can be replaced with a higher limit.

\begin{table*}[t]
    \caption{The rules to restrict queries.}
    \centering\small
    \begin{threeparttable}
    \begin{tabular}{@{}rllll@{}}
    \toprule
    & \textbf{Clause} & \textbf{Source} & \textbf{Target} & \textbf{Example} \\
    \midrule
    1 & JOIN & LEFT JOIN & INNER JOIN & SELECT * FROM t0 \colorbox{shallowred}{\sout{LEFT}} \colorbox{shallowgreen}{INNER} JOIN t1 ON ...; \\
    2 & JOIN & RIGHT JOIN & INNER JOIN & SELECT * FROM t0 \colorbox{shallowred}{\sout{RIGHT}} \colorbox{shallowgreen}{INNER} JOIN t1 ON ...; \\
    3 & JOIN & FULL JOIN & LEFT JOIN & SELECT * FROM t0 \colorbox{shallowred}{\sout{FULL}} \colorbox{shallowgreen}{LEFT} JOIN t1 ON ...; \\
    4 & JOIN & FULL JOIN & RIGHT JOIN & SELECT * FROM t0 \colorbox{shallowred}{\sout{FULL}} \colorbox{shallowgreen}{RIGHT} JOIN t1 ON ...; \\
    5\tnote{\dag} & JOIN & CROSS JOIN & FULL JOIN & SELECT * FROM t0 \colorbox{shallowred}{\sout{CROSS}} \colorbox{shallowgreen}{FULL} JOIN t1; \\
    6 & SELECT & ALL & DISTINCT & SELECT \colorbox{shallowred}{\sout{ALL}} \colorbox{shallowgreen}{DISTINCT} * FROM t0; \\
    7 & GROUP BY & <Empty> & <Predicate> & SELECT * FROM t0 \colorbox{shallowgreen}{GROUP BY c0}; \\
    8 & HAVING & <Empty> & <Predicate> & SELECT * FROM t0 GROUP BY c0 \colorbox{shallowgreen}{HAVING c0>0};\\
    9 & WHERE & <Empty> & <Predicate> & SELECT * FROM t0 \colorbox{shallowgreen}{WHERE c0>0};\\
    10 & WHERE & <Predicate> & <Predicate> AND <Predicate> & SELECT * FROM t0 WHERE c0>0 \colorbox{shallowgreen}{AND c0!=8}; \\
    11 & WHERE & <Predicate> OR <Predicate> & <Predicate> & SELECT * FROM t0 WHERE c0>0 \colorbox{shallowred}{\sout{OR c0!=8}};\\
    12 & LIMIT & <Natural number> & <Natural number> - <Natural number> & SELECT * FROM t0 LIMIT \colorbox{shallowred}{\sout{10}} \colorbox{shallowgreen}{5}; \\
    \bottomrule
    \end{tabular}
    \begin{tablenotes}
        \item[\dag] Rule 5 holds when both tables are not empty.
    \end{tablenotes}
    \end{threeparttable}
    \label{tab:rules}
\end{table*}

\subsection{Checking for Structural Similarity}

\begin{figure}\footnotesize
\begin{lstlisting}[caption={An example of query plans that are not structurally similar, which is why we exclude them for testing.},captionpos=t, label=lst:struequ, escapeinside=&&]
CREATE TABLE t0 (c0 INT);
CREATE TABLE t1 (c0 INT, c1 INT);
INSERT INTO t1 VALUES(1,2), (3,4), (5,6), (NULL, NULL);
INSERT INTO t0 VALUES(1), (2);
ANALYZE t0;
ANALYZE t1;

EXPLAIN SELECT * FROM t0 &\underline{\textbf{FULL JOIN}}& t1 ON t1.c1 IN (t1.c1) WHERE CASE WHEN t1.rowid > 2 THEN false ELSE t1.c1=1 END; -- estimated rows: 2
EXPLAIN SELECT * FROM t0 &\underline{\textbf{RIGHT JOIN}}& t1 ON t1.c1 IN (t1.c1) WHERE CASE WHEN t1.rowid > 2 THEN false ELSE t1.c1=1 END; -- estimated rows: 3
--------------------------------------------------
&$\bullet$& filter                 &$\bullet$& cross join(right)
| estimated row:2        | estimated row:3
|-&$\bullet$& cross join(full)     |-&$\bullet$& scan (t0)
   | estimated row:6     |    estimated row:2 
   |-&$\bullet$& scan (t1)         |-&$\bullet$& filter
   |    estimated row:4     | estimated row:1 
   |-&$\bullet$& scan (t0)            |-&$\bullet$& scan (t1)
        estimated row:2          estimated row:4     
\end{lstlisting}
\end{figure}

Even for similar queries, DBMSs may create significantly different query plans. In such cases, the estimated cardinalities might be calculated in different ways, and thus result in false alarms.
\Autoref{lst:struequ} shows an example of this problem. The only difference between the two queries in lines 8--9 is that \lstinline{FULL JOIN} is used in the first query and \lstinline{RIGHT JOIN} is used in the second query. Their estimated cardinalities are 2 and 3 respectively. Based on rule 4 alone, this discrepancy would constitute a performance issue; however, consider the query plans in \Autoref{lst:struequ}.
In lines 11--18, the left part is the query plan of the first query, and the right part is that of the second query. For the first query with \lstinline{FULL JOIN}, the sequence of operations is \lstinline{filter, cross join, scan, scan} in which the operation \lstinline{filter} is applied \emph{after} the operation \lstinline{cross join}. For the second query with \lstinline{RIGHT JOIN}, the sequence of operations is \lstinline{cross join, scan, filter, scan}, in which the operation \lstinline{filter} is applied \emph{before} the operation \lstinline{cross join}. The difference is due to a SQL optimization mechanism called \emph{predicate pushdown}~\cite{levy1994query}, which moves a filter to be executed before joining two tables and is applied in the second query. The predicate pushdown does not affect the final result, but can reduce the estimated cardinalities to be joined in the operation \lstinline{RIGHT JOIN}, which is more efficient. However, because the predicate is pushed down, the structure of the query plans changes. Therefore, the estimated cardinalities are calculated in a different manner than that of the first query, and the developers consider the estimated cardinalities of both query plans as incomparable.\footnote{\url{https://github.com/cockroachdb/cockroach/issues/89060}} In \Autoref{lst:struequ}, the estimated cardinalities of both operations \lstinline{FULL JOIN} and \lstinline{RIGHT JOIN} are calculated as the sum of the estimated cardinalities in the last step and the operation \lstinline{filter} is calculated as one-third of the estimated cardinalities in the last step. The estimated cardinality of the first query plan is calculated by $(2 + 4) / 3 = 2$, while that of the second query plan is calculated by $4 / 3 + 2 =3$. 

We identify comparable estimated cardinalities by checking for \emph{structural similarity}. In a query plan $P$, the operation of a node $N$ is denoted as $O_N$. Suppose $N$ has $k$ children, denoted as $C_1, C_2, \ldots, C_k$. The node's flattened operation sequence is an array obtained by concatenating the flattened child nodes: $\text{flatten}(N) = [O_N, \text{flatten}(C_1) , \text{flatten}(C_2) \ldots \text{flatten}(C_k)]$. The flattening of the root node is also denoted as $\text{flatten}(P)$. For a pair of query plans $P_a$ and $P_b$, we define both are \emph{structurally similar} only if $ED(\text{flatten}(P_a), \text{flatten}(P_b)) <= 1$, in which $ED$ represents the edit distance~\cite{navarro2001guided}, a common way of quantifying the dissimilarity of two strings, of both operation sequences. For example, in the query plans of \Autoref{fig:approach}, the sequences of operations are \lstinline{[cross join, scan, scan]} and \lstinline{[cross join, filter, scan, scan]}. The first sequence can be edited to the second sequence by inserting a \lstinline{filter} only, and \emph{vice versa}. Both query plans are structurally similar and we validate the \property property. If they are not structurally similar, we continue testing with a new query. The calculation is based on sequences rather than trees, because the computation of the edit distance of trees was shown to be NP-hard~\cite{treeeditdistance}.

\subsection{Validating Cardinality Estimation}\label{subsec:validating}
Finally, we validate the \property property on the estimated cardinalities extracted from the query plans. If the estimated cardinality of the original query is lower than that of the more restrictive query, we report the query pair as an issue. In \Autoref{fig:approach}, the estimated cardinality of the original query is 20, which is lower than that of the other restricted query, which is 60. This indicates an unexpected result, and we report both queries to developers. Recall that we deem the estimated cardinality in the root operation of the query plan as the estimated cardinality of the query and ignore the estimated cardinalities of other operations.

\section{Evaluation}

To evaluate the effectiveness and efficiency of \method in finding performance issues through estimated cardinalities, we implemented \method in \sqlancer,\footnote{\url{https://github.com/sqlancer/sqlancer/releases/tag/v2.0.0}} which is an automated testing tool for DBMSs, and, based on our prototype \tool, we sought to answer the following questions:
\begin{description}
    \item[\textbf{Q.1 Effectiveness.}] Can \method identify previously unknown issues?
    \item[\textbf{Q.2 Historic Bugs.}] Can \method identify historic performance issues?
    \item[\textbf{Q.3 Efficiency.}] How does CERT compare to the state-of-the-art approach in terms of accuracy and efficiency?
    \item[\textbf{Q.4 Sensitivity.}] Which rules proposed in \Autoref{tab:rules} contribute to finding issues? Do DBMSs adhere to the \property property in practice?
\end{description}

\paragraph{Implementation.}  We reused \sqlancer and its implementation of a rule-based random generation method to generate queries and database states. For each table, we generated 100 \lstinline{INSERT} statements and ensured that every table contained at least one row. Then, the generated queries and database states are passed to \method for validating the \property property. The core logic of \method is implemented in only around 200 lines of Java code for each DBMS, suggesting that its low implementation effort might make the approach widely applicable. We used pattern matching to extract operations and estimated cardinalities, and implemented the \emph{Dynamic Programming} (DP) algorithm~\cite{dp} to calculate the structural similarity.

\paragraph{Tested DBMSs.} We tested the same DBMSs, MySQL, TiDB, and CockroachDB as we studied in \Autoref{sec:study}. For Q1, Q3, and Q4, we used the latest available development versions (MySQL: 8.0.31, TiDB: 6.4.0, CockroachDB: 22.2.0).  For Q3, in an attempt of a fairer comparison to \amoeba, we chose the historical version of CockroachDB 20.2.19, which is the version that \amoeba~\cite{liu2022automatic} tested.

\paragraph{Baselines.} To the best of our knowledge, no existing work can be applied to specifically test cardinality estimation. The most closely related work is \amoeba, which finds performance issues in query optimizers. We did not consider \apollo~\cite{jung2019apollo}, because it finds only performance regressions. Ensuring a fair comparison with \amoeba is challenging, as the approaches are not directly comparable.
\amoeba validates that semantic-equivalent queries exhibit similar performance characteristics, while \method validates the \property property.
Furthermore, both tools support a different set of DBMSs; \amoeba supports CockroachDB and PostgreSQL, while \method supports CockroachDB, TiDB, and MySQL. Thus, we performed the comparison in Q3 using only CockroachDB, which is supported by both tools.

\paragraph{Experimental infrastructure.} We conducted all experiments on a desktop computer with an Intel(R) Core(TM) i7-9700 processor that has 8 physical cores clocked at 3.00GHz. Our test machine uses Ubuntu 20.04 with 8 GB of RAM, and a maximum utilization of 8 cores. We run all experiments 10 runs for statistical significance.

\subsection*{Q.1 Effectiveness}\label{sssec:Q1}

\begin{table*}
    \centering
    \caption{The issues found by \method. The red strike-through text is removed, while the content in green color is added.}
    \begin{tabular}{@{}lllrll@{}}
        \toprule
        \textbf{DBMS} & \textbf{Bug ID} & \textbf{Version} & \textbf{Rules} & \textbf{Modifications to the query} & \textbf{Status} \\
       \midrule
MySQL         & 108833    & 8.0.31   & 9      & ...\colorbox{shallowgreen}{WHERE t0.c0 > t0.c1}...                                          & Verified \\
MySQL         & 108851    & 8.0.31   & 9      & ...\colorbox{shallowgreen}{WHERE t1.c1 BETWEEN (SELECT 1 WHERE FALSE) AND (t1.c0)}...       & Verified \\
MySQL         & 108852    & 8.0.31   & 6      & ...\colorbox{shallowgreen}{DISTINCT}...                                                     & Verified \\
TiDB          & 38319     & 51a6684f & 11     & ...WHERE (TRUE)\colorbox{shallowred}{\sout{OR(TO\_BASE64(t0.c0))}}...                       & Confirmed \\
TiDB          & 38747     & 3ef8352a & 7      &  ...\colorbox{shallowgreen}{GROUP BY t0.c0}...                                              & Confirmed \\
TiDB          & 38479     & 3ef8352a & 3 \& 5 & ...\colorbox{shallowred}{\sout{CROSS}}\colorbox{shallowgreen}{LEFT} JOIN...                 & Confirmed \\
TiDB          & 38482     & 3ef8352a & 8      & ...\colorbox{shallowgreen}{HAVING (t1.c0)REGEXP(NULL)}...                                   & Confirmed \\
TiDB          & 38665     & 6c55faf0 & 2      & ...\colorbox{shallowred}{\sout{RIGHT}}\colorbox{shallowgreen}{INNER} JOIN...                & Confirmed \\
TiDB          & 38721     & 6c55faf0 & 9      & ...\colorbox{shallowgreen}{WHERE v0.c2}...                                                  & Confirmed \\
CockroachDB   & 88455	  & 7cde315d & 1      & ...\colorbox{shallowred}{\sout{LEFT}}\colorbox{shallowgreen}{INNER} JOIN...                 & Fixed \\
CockroachDB   & 89161	  & f188d21d & 11     & ...WHERE (t0.c0 IS NOT NULL) \colorbox{shallowred}{\sout{OR (1 < ALL (t0.c0 \&  t0.c0))}}...& Fixed (Known) \\
CockroachDB   & 89462	  & 81586f62 & 8      & ...\colorbox{shallowgreen}{HAVING (t1.c0 ::CHAR) = 'a'}...                                  & Backlogged \\
CockroachDB   & 90113	  & fbfb71b9 & 2      & ...\colorbox{shallowred}{\sout{RIGHT}}\colorbox{shallowgreen}{INNER} JOIN...                & Backlogged \\

        \bottomrule
    \end{tabular}

    \label{tab:bugs}
\end{table*}

\paragraph{Method.} We ran \tool to find performance issues. Each automatically generated issue report usually includes many SQL statements, making it challenging for developers to analyze the root reason for the issue. To alleviate this problem and better demonstrate the underlying reasons for these issues, we adopted delta debugging~\cite{zeller2009programs} to minimize test cases before reporting them to developers. The steps to minimize the test case are 1) incrementally removing some of the SQL statements in the test case and 2) ensuring that the \property property is still violated. After submitting the issue reports with minimized test cases to developers, we submitted follow-up issues only if we believed them to be unique, such as those identified by different rules than previous issues, to avoid duplicate issues. MySQL has its own issue-tracking system, and developers add a label \emph{Verified} for the issues that they have confirmed. TiDB and CockroachDB use GitHub's issue tracker. TiDB's developers assign labels, such as affected versions and modules; we considered the issue as \emph{Confirmed} after such a label was assigned. CockroachDB's developers typically directly replied whether they planned on fixing the issue which we consider \emph{Fixed} or whether the issue was considered a false alarm. In some cases, they added a \emph{Backlogged} label to indicate that they would investigate this issue in the future. For all DBMSs, based on historic reports, we observed that, typically, developers directly reject duplicate issue reports.

\paragraph{Results.} \Autoref{tab:bugs} shows the unique issues that \method found in three tested DBMSs. The \emph{Bug ID} column shows the bug id in respective bug trackers. The \emph{Version} column shows the versions or git commits of the DBMSs in which we found corresponding issues. The \emph{Rules} column shows which rules identified this issue. In total, we have reported \numbugs unique issues on unexpected estimated cardinalities. \numconfirmedbugs issues have been confirmed by developers in three days, \numfixedbugs issues have been fixed in one week, and 2 issues were backlogged. No false alarm was generated. We speculate that many confirmed bugs remain unfixed, because 1) fixing performance issues requires comprehensive consideration which usually consumes much time, and 2) performance issues might have lower priority than other issues, such as correctness bugs, which cause a query to compute an incorrect result. Among all \numbugs unique issues, the only known issue that we found in CockroachDB had been backlogged for around 10 months since it was first found, and our test case clearly demonstrated the root reason for the issue, which allowed developers to quickly fix it. Apart from the reported issues, \method continuously generates more than ten issue reports per minute. We did not report the additional bug-inducing test cases to the developers to avoid burdening them, because deciding their uniqueness would be challenging. Therefore, we believe that \method could help identify additional performance issues in the future. Overall, all issues were exposed in various SQL clauses and predicates, which may imply no common issues across tested DBMSs. We give two examples of minimized test cases to explain the issues we found as follows.

\begin{figure}
\begin{lstlisting}[caption={Rule 6, which identified \#108852 in MySQL, replaces ALL with DISTINCT.},captionpos=t, label=lst:case1, escapeinside=@@]
CREATE TABLE t0(c0 INT, c1 INT UNIQUE) ;
INSERT INTO t0 VALUES(-1, NULL),(1, 2),(NULL, NULL),(3, 4);
ANALYZE TABLE t0 UPDATE HISTOGRAM ON c0, c1;

EXPLAIN SELECT @\underline{ALL}@ t0.c0 FROM t0 WHERE t0.c1; -- estimated rows: 3
EXPLAIN SELECT @\underline{DISTINCT}@ t0.c0 FROM t0 WHERE t0.c1; -- estimated rows: 4
\end{lstlisting}
\end{figure}

\paragraph{Issue \#108852 identified by rule 6.}
\Autoref{lst:case1} shows a test case exposing issue \#108852 in MySQL. Rule 6, which replaces \lstinline{ALL} with \lstinline{DISTINCT} in \Autoref{tab:rules}, exposed this issue. In \Autoref{lst:case1}, the first query in line 5 fetches the rows including duplicate rows, while the second query in line 6 excludes duplicate rows, so the cardinality of the second query should be no more than that of the first query. However, the estimated cardinality of the second query is greater than that of the first query, which is unexpected. Suppose a query $q$ with \lstinline{ALL} is a subquery of another query $Q$ with \lstinline{DISTINCT}, this issue affects whether \lstinline{DISTINCT} should be pushed down to the execution of $q$ for an efficient query plan that aims to retrieve the fewest rows from $q$. This issue was confirmed by the MySQL developers already three hours after we reported it. 

\begin{figure}
\begin{lstlisting}[caption={Rule 11, which identified \#89161 in CockroachDB, removes either operand of an OR expression.},captionpos=t, label=lst:case2, escapeinside=@@]
CREATE TABLE t0 (c0 INT);
INSERT INTO t0 VALUES (1), (2), (3), (4), (5), (6), (7), (8), (9), (10);
ANALYZE t0;

EXPLAIN SELECT t0.c0 FROM t0 WHERE @\underline{(t0.c0 IS NOT NULL) OR (1 < ALL (t0.c0, t0.c0))}@; -- estimated rows: 3
EXPLAIN SELECT t0.c0 FROM t0 WHERE @\underline{(t0.c0 IS NOT NULL)}@; -- estimated rows: 10
\end{lstlisting}
\end{figure}

\paragraph{Issue \#89161 identified by rule 11.} 
\Autoref{lst:case2} shows a test case exposing issue \#89161 in CockroachDB by rule 11, which removes either operand of an OR expression. The predicate \lstinline{(t0.c0 IS NOT NULL)} in the \lstinline{WHERE} clause of the second query should fetch no more rows than the predicate \lstinline{(t0.c0 IS NOT NULL) OR (1 < ALL (t0.c0, t0.c0))} of the first query. However, the estimated cardinality of the second query is greater than that of the first query, which is unexpected. This issue was caused by a buggy logic to handle the \lstinline{OR} clause. In CockroachDB, given predicates \lstinline{A} and \lstinline{B}, the estimated cardinality of predicate \lstinline{A OR B} is calculated by: $P(A \; OR \; B) = P(A) + P(B) - P(A \; AND \; B)$. However, when \lstinline{A} and \lstinline{B} depend on the same table or column, the estimated cardinality is unexpected. We found this issue by rule 11 in \Autoref{tab:rules}. Although this issue was known, it had been backlogged for around 10 months since it was first found. When we reported our test case, the developer opened a pull request in their git repository to fix it after three days.

\begin{figure}
\begin{lstlisting}[caption={The performance improvement by fixing our found issues \#88455 and \#89161.},captionpos=t, label=lst:case3, escapeinside=@@]
CREATE TABLE t0 (c0 INT);
CREATE TABLE t1 (c0 INT);
CREATE TABLE t2 (c0 INT);
INSERT INTO t0 SELECT * FROM generate_series(1,1000);
INSERT INTO t1 SELECT * FROM generate_series(1001,2000);  
INSERT INTO t2 SELECT * FROM generate_series(1,333100);
ANALYZE t0; 
ANALYZE t1; 
ANALYZE t2;

SELECT COUNT(*) FROM t0 LEFT OUTER JOIN t1 ON t0.c0<1 OR t0.c0>1 FULL JOIN t2 ON t0.c0=t2.c0; -- 399ms @$\rightarrow$@ 321ms
SELECT COUNT(*) FROM t0 LEFT JOIN t1 ON t0.c0>0 WHERE (t0.c0 IS NOT NULL) OR (1 < ALL(t0.c0, t0.c0)); -- 131ms @$\rightarrow$@ 109ms

\end{lstlisting}
\end{figure}

\paragraph{Performance analysis.} To investigate the extent to which the issues we found affect performance, we evaluated the query performance of the fixed issues \#88455 and \#89161 on a test case as shown in \Autoref{lst:case3} that involves joining multiple tables. We could not consider unfixed issues, as it would be unclear how to determine the potential speedup. We executed both queries in lines 12 and 13 before and after the fixes of \#88455 and \#89161 respectively. After executing either query ten times, we found that the fixes improved the performance by an average of 20\% and 17\%, respectively. This improvement is due to the more accurate estimated cardinality which allows for more optimal joining orders.

\result{Using \method, we have found \numbugs unique issues in MySQL, TiDB, and CockroachDB. The fixes improve query performance by 19\% on average.}

\subsection*{Q.2 Historic Bugs}

\paragraph{Method.} To evaluate whether \property is sufficiently general to identify previous performance issues that we identified in \Autoref{tab:study}, we attempted using \method to identify all three performance issues whose fixes changed the estimated cardinalities, namely issues \#61631, \#56714, and \#9067. Specifically, based on the queries in the issue reports, we followed step \textcircled{2} in \Autoref{fig:approach} to randomly construct 10,000 pairs of queries. Then, we checked whether any pair violated the \property before the fix, and adhered to the \property after the fix. If so and both query plans are structurally similar, we concluded that \property could have identified the performance issue.

\begin{figure}
\begin{lstlisting}[caption={Issue \#56714 violates the \property.},captionpos=t, label=lst:case4, escapeinside=@@]
...
EXPLAIN SELECT MAX(a) FROM test; -- estimated rows: 1
EXPLAIN SELECT MAX(a) FROM test WHERE b=0; -- estimated rows: 500190
\end{lstlisting}
\end{figure}

\paragraph{Results.} All three previous performance issues caused by cardinality estimation can be found by \method. For example, considering the performance issue \#56714 in \Autoref{lst:prebug}, \Autoref{lst:case4} shows the pair of queries that \method produces to identify the performance issue. The second query has an additional \lstinline{WHERE} clause compared to the first query, so the estimated cardinality of the second query should be no more than that of the first query. However, due to incorrect usage of the index in column \lstinline{b}, the second query scans all rows and has a higher estimated cardinality. After the fix, the estimated cardinality of the second query decreases to 1.

\result{The \property can identify historical performance issues caused by cardinality estimation.}

\subsection*{Q.3 Efficiency}\label{sssec:Q3}

\paragraph{Accuracy.} We evaluated whether \method has higher accuracy in confirmed issues than \amoeba. We evaluated this aspect based on the observation that around five in six reported bugs by \amoeba were false alarms. A high rate of false alarms significantly limits the applicability of an automated testing technique. Recall that it is challenging to make a fair comparison as \method and \amoeba find different kinds of issues affecting performance. 

\paragraph{Results.} \Autoref{tab:confirmedrate} shows the number of all and confirmed/fixed unique performance issues found by \method and \amoeba. The authors of \amoeba reported 25 issues in CockroachDB, but only \amoebaconfirmed issues (\amoebaconfirmedrate accuracy) were confirmed or fixed by developers. In comparison, for \method, 50\% of issues in CockroachDB and 100\% of issues in MySQL and TiDB were confirmed or fixed. For CockroachDB, \method found fewer performance issues than \amoeba, because we did not report all found issues to avoid duplicate reports. Overall, these results suggest that \method has high accuracy and is a practical technique for finding relevant performance issues. Despite these promising results, on a conceptual level, similar to \amoeba, we cannot ensure that the performance issues would be considered as such by the developers.

\begin{table}
    \centering\small
    \caption{The number of all (\textbf{All}) and confirmed or fixed (\textbf{C/F}) unique performance issues.}
    \begin{tabular}{@{}l@{}rrrrrr@{}r@{}}
        \toprule
                    & \multicolumn{3}{c}{\textbf{\method}} & \multicolumn{3}{c}{\textbf{\amoeba}} & \\
                    \cmidrule(lr){2-4}
                    \cmidrule(lr){5-7}
       \textbf{DBMS} & \textbf{All} & \textbf{C/F} & \textbf{\%} & \textbf{All} & \textbf{C/F} & \textbf{\%} \\
       \midrule
       MySQL & 3 & 3 & 100\% & - & - & - \\
       TiDB & 6 & 6 & 100\% & - & - & - \\
       CockroachDB & 4 & 2 & 50\% & 25 & 6 & 24\% \\
        \bottomrule
       \textbf{Sum:} & 13 & 11 & \confirmedrate & 25 & 6 & \amoebaconfirmedrate \\
    \end{tabular}
    \label{tab:confirmedrate}
\end{table}

\paragraph{Throughput.} We evaluated whether \method has a higher testing throughput than \amoeba. State-of-the-art benchmarks and approaches, such as TPC-H~\cite{tpch}, \amoeba~\cite{liu2022automatic}, and \apollo~\cite{jung2019apollo} execute queries, which results in relatively low throughput. Therefore, we expect that \method will validate more queries per second. To evaluate this, we determined the average number of test cases per second processed by \tool and \amoeba in one hour. 

\paragraph{Results.} In CockroachDB, on average across 10 runs and one hour, \method validates 714.54 test cases while \amoeba exercises 1.85 test cases per second. This suggests a \speedfactor performance improvement over \amoeba. Recall that the throughput results are not directly comparable, as different approaches can find different kinds of issues. In addition, \method is immune to performance fluctuation, because the estimated cardinalities are not affected by execution time. Therefore, \method yields the same results in different hardware and network environments.

\result{\tool validates \speedfactor more test cases than \amoeba across one hour and 10 runs on average, and is more than twice as accurate as \amoeba.}

\subsection*{Q.4 Sensitivity}\label{sssec:Q4}

\paragraph{Sensitivity of rules.} We evaluated which rules presented in \Autoref{tab:rules} contribute to finding the issues in \Autoref{tab:bugs}. Specifically, we recorded which rules were applied to the bug-inducing test cases that we reported. We considered reported bug-inducing test cases, rather than all test cases---recall that \tool still reports violations when being run---as we expect the reported issues to be unique based on the developers' verdicts. \Autoref{tab:bugs} shows the rules applied to the corresponding bug-inducing test cases. Overall, 9 out of 12 rules have found at least one issue. Rule 9, which adds a predicate to \lstinline{WHERE} clause, found the most issues, namely three. We believe that this is because the predicates in \lstinline{WHERE} clause can vary significantly and thus be diverse and have a higher possibility of exposing issues. No issue was found by rules 4, 10, and 12. Rule 4 applies to the \lstinline{JOIN} clause in which other rules found several issues. Similarly, we believe that rule 10, which restricts a \lstinline{WHERE} clause by an \lstinline{AND} operator would find issues after the issues found by other rules applied to \lstinline{WHERE} clause are fixed. Rule 12 applies to the \lstinline{LIMIT} clause, which simply returns up to as many rows as specified in its argument. We explain that the simplicity of \lstinline{LIMIT} explains that we have found no issues in its handling. 

\begin{table}[]
    \centering\small
    \caption{The average number of all queries (\textbf{Queries}), the queries that violate \property (\textbf{Violations}), and the geometric mean of percentage (\%) of queries that violate the property across 10 runs and one hour.}
    \begin{tabular}{@{}lrrr@{}}
        \toprule
        \textbf{DBMS} & \textbf{Queries (\#)} & \textbf{Violations (\#)} & \textbf{Violations (\%)} \\
        \midrule
        MySQL & 6,371,222 & 30,841 & 0.28\% \\
        TiDB & 2,895,203 & 8,108 & 0.27\% \\
        CockroachDB & 1,306,807 & 661 & 0.05\% \\
        \bottomrule
        & \multicolumn{2}{r}{} & \textbf{Average:} \violatedpercentage
    \end{tabular}
    \label{tab:violation}
\end{table}

\paragraph{Sensitivity of \property.} We expect that any violation of the \property property indicates a potential issue. To more thoroughly assess our hypothesis, we examined how many queries among all tested queries violate the \property property. If only a small portion of queries violate it, DBMSs are likely to adhere to the \property property, and any violation warrants further investigation. Otherwise, the property may be not meaningful for developers. \Autoref{tab:violation} shows the average number of all queries, the queries that violate \property property, and the geometric mean of the percentage of queries that violate the property across 10 runs and one hour. Overall, \violatedpercentage of all generated queries violate the \property property. All three DBMSs, MySQL, TiDB, and CockroachDB, exhibit a similar rate of \property violations. The results demonstrate that DBMSs typically comply with the \property property, as more than 99.5\% of generated queries do not violate it.
\section{Discussion}
We discuss some key considerations on the design of \method, its characteristics, as well as the evaluation's results.

\paragraph{Evaluating performance gains.} It is challenging to measure the overall performance gain that fixing the issues reported by \method could achieve in practice. One issue is that measuring the overall performance impact might be misleading, because many other components in query optimizers can affect performance as well. For example, research on the Join Order Benchmark~\cite{leis2015good} demonstrated that a worse cardinality estimator might lead to better performance due to the issues in other components. In addition, \numconfirmedbugs issues were confirmed, but remained unfixed. Since we lack domain knowledge to address the underlying issues, we cannot determine the performance gains that fixing these issues would cause.

\sloppy{}
\paragraph{Generality.} 
While we focused on important SQL clauses in this work, the 12 rules shown in \Autoref{tab:rules} are not comprehensive. They could be extended to cover various other features, such as window functions. Additionally, the \property property might be applicable also to other kinds of data models than the traditional relational data model. For example, Neo4J, a graph DBMS, also uses a concept similar to query plans---termed execution plans---and cardinality estimation\footnote{\url{https://neo4j.com/docs/cypher-manual/current/execution-plans/\#execution-plan-introduction}} (\emph{EstimatedRows} field in execution plans). Its query optimization also depends on cardinality estimation, which we expect to comply with the \property property. More work is required to explore \property in other DBMSs in the future. 

\paragraph{Threats to validity.} Our evaluation results face potential threats to validity.
One concern is internal validity, that is, the degree to which our results minimize systematic error. \method validates test cases that are randomly generated by \sqlancer. The randomness process may limit the reproducibility of our results. To mitigate the risk, we repeated all experiments 10 times to gain statistical significance. Another concern is external validity, that is, the degree to which our results can be generalized to and across other DBMSs. We selected various types of DBMSs including different purposes (community-developed: MySQL and company-backed: TiDB and CockroachDB) and languages (C/C++: MySQL and Go: TiDB and CockroachDB). These DBMSs have been widely used in prior research~\cite{liang:sqlright, Rigger2020PQS, Rigger2020TLP}. Given that DBMSs provide similar functionality and features, we are confident that our results generalize to many DBMSs. The last concern is construct validity, that is, the degree to which our evaluation accurately assesses what the results are supposed to. \method found \numbugs unique performance issues, but only \numfixedbugs issues had been fixed posing the threat that developers might not fix them in the future or might deem them less important. To address this threat, we communicated with the TiDB developers, who informed us that they plan to fix the issues and indicated an interest in using \tool.

\section{Related Work}

\paragraph{Automatically testing for performance issues.} 
The most related strand of research is on finding performance issues in DBMSs automatically. Jung \etal proposed \apollo~\cite{jung2019apollo}, which compares the execution times of a query on two versions of a database system to find performance regression bugs. Liu \etal proposed \amoeba~\cite{liu2022automatic}, which compares the execution time of a semantically-equivalent pair of queries to identify an unexpected slowdown.
In contrast, \method specifically tests cardinality estimation, which is most critical for query optimization~\cite{leis2015good}.
Thus, we believe that issues found by \method might be most relevant for DBMSs' performance.
Unlike these works, \method avoids executing queries by inspecting only query plans, allowing for higher throughput and making the approach robust against unanticipated performance fluctuations. 

\paragraph{Performance benchmarking.} Benchmarking DBMSs is a common practice to identify performance regressions, and to continuously improve the DBMSs' performance on a set of benchmarks. \emph{TPC-H}~\cite{tpch} and \emph{TPC-DS}~\cite{tpcds} are the most popular benchmarks and are considered the industry standard. Boncz \etal studied and identified 28 ``chokepoints'' (\emph{i.e.}, optimization challenges) of the TPC-H benchmark~\cite{boncz2013tpc}. Poess \etal modified and analyzed the TPC-DS benchmark~\cite{Poess2017} to measure SQL-based big data systems. Karimov \etal proposed a benchmark for stream-data-processing systems~\cite{Karimov2018}. 
Boncz \etal proposed an improved TPC-H benchmark,\emph{JCC-H}~\cite{Boncz2018}, which introduces \emph{Join-Crossing-Correlations} (JCC) to evaluate the scenarios where data in one table can affect the behaviors of operations involving data in other tables. Leis \etal proposed the \emph{Join Order Benchmark (JOB)}~\cite{leis2015good}, which uses more complex join orders. 
Raasveldt \etal described common pitfalls when benchmarking DBMSs and demonstrated how they can affect a DBMS's performance~\cite{Raasveldt2018}.
\method is complementary to benchmarking; while benchmarking focuses on workloads deemed relevant for users, \method can find performance issues through the lens of cardinality estimation even on previously unseen workloads.

\paragraph{DBMS testing.} Besides testing DBMSs' performance, automated testing approaches have been proposed to find other types of bugs. Many fuzzing works on finding security-relevant bugs, such as memory errors, were proposed. SQLSmith~\cite{sqlsmith}, Griffin~\cite{fu2022griffin}, DynSQL~\cite{jiangdynsql}, and ADUSA~\cite{liu2022automatic} used grammar-based methods to generate test cases for detecting memory errors. Squirrel~\cite{zhong2020squirrel}, inspired by grey-box fuzzers such as AFL~\cite{afl}, used code coverage as guidance to find memory errors. 
Various approaches have been proposed to find logic bugs in DBMSs. The PQS~\cite{Rigger2020PQS}, NoREC~\cite{Rigger2020NoREC}, and TLP~\cite{Rigger2020TLP} oracles detect logic bugs in the implementation of \lstinline{SELECT} statements. DQE~\cite{song2023testing} detects logic bugs in \lstinline{UPDATE} and \lstinline{INSERT} statements. Transactional properties have also been tested. Cobra~\cite{tan2020cobra} used formal methods to verify the serializability of executions in key-value stores. Troc~\cite{dou2023detecting} defines an oracle to detect isolation bugs in transactional procedures.
Similar to testing approaches that find logic bugs and test properties, our core contribution is a new test oracle, which, however, finds performance issues that violate the \property property.

\paragraph{Cardinality estimation.} Various cardinality-estimation techniques have been proposed. Han \etal~\cite{han2021cardinality} comprehensively evaluated various algorithms for cardinality estimation, of which we describe important ones below. PostgreSQL~\cite{postgresql} and MultiHist~\cite{poosala1997selectivity} applied one-dimensional and multi-dimensional histograms to estimate cardinality. Similarly,  UniSample~\cite{leis2017cardinality, zhao2018random} and WJSample~\cite{li2016wander} used sampling-based methods to estimate cardinality. Apart from these traditional approaches, machine learning-based methods have gained attention recently. MSCN~\cite{kipf2018learned, naru}, LW-XGB~\cite{dutt2019selectivity}, and UAE-Q~\cite{wu2021unified} used deep neural networks, classic lightweight regression models, and deep auto-regression models to learn to map each query directly to its estimated cardinality. In addition,  NeuroCard~\cite{neurocard}, BayesCard~\cite{wu2020bayescard}, DeepDB~\cite{deepdb}, and FSPN~\cite{wu2020fspn, zhu2020flat} utilized deep auto-regression models and three probabilistic graphical models BN, SPN, and FSPN to predict the data distribution for cardinality estimation. \method is a black-box technique that could, in principle, be applied to any cardinality estimator. However, we believe that finding relevant issues in learning-based estimators that can be fixed might be challenging.

\paragraph{Metamorphic testing.} At a conceptual level, \method can be classified as a metamorphic testing approach~\cite{chen2002metamorphic, chen2018metamorphic}, which is a method to generate both test cases and validate results. Metamorphic testing uses an input $I$ to a system and its output $O$ to derive a new input $I'$ (and output $O'$), for which a test oracle can be provided that checks whether a so-called \emph{Metamorphic Relation} holds between $O$ and $O'$. For \method, $I$ corresponds to the original query, $O$ is the estimated cardinality, $I'$ is the restricted query, and $O'$ is its estimated cardinality.
The metamorphic relation that we validate is the \property property.
Metamorphic testing has been applied successfully in various domains~\cite{segura2018metamorphic, chen2018metamorphic}.
TLP~\cite{Rigger2020TLP} and NoREC~\cite{Rigger2020NoREC} are other metamorphic testing approaches that were proposed for testing DBMSs, as discussed above.
\section{Conclusion}
We have presented \method, a novel technique for finding performance issues through the lens of cardinality estimation in DBMSs.
Our key idea is to, given a query, derive a more restrictive query, and validate that the restriction is reflected by the DBMSs' estimated cardinalities; we refer to this property as \property.
Specifically, we validate that the original query's cardinality is at least as high as the more restrictive query. Our evaluation has demonstrated that this technique is effective.
Of the \numbugs unique issues that we reported, \numfixedbugs issues were fixed, \numconfirmedbugs issues were confirmed, and 2 issues require further investigation.
The fixes improved query performance by 19\% on average.
Unlike other testing approaches for performance issues, \method avoids executing queries, achieving a speedup of \speedfactor compared to the state of the art.
Finally, it is readily applicable.
DBMSs expose estimated cardinalities as part of query plans to users; thus, \method is a black-box technique that is applicable without modifications, even if the DBMSs' source code is inaccessible to the testers.
Furthermore, since no queries are executed, \method is resistant to performance fluctuations.
Overall, we believe that \method is a useful technique for DBMS developers and testers and hope that the technique will be widely adopted in practice.

\section{Data Availability}
Our implementation and experimental data are publicly available at \url{https://zenodo.org/records/10476847}.

\begin{acks}
This research is supported by the National Research Foundation, Singapore, and Cyber Security Agency of Singapore under its National Cybersecurity R\&D Programme (Fuzz Testing <NRF-NCR25-Fuzz-0001>). Any opinions, findings and conclusions, or recommendations expressed in this material are those of the author(s) and do not reflect the views of National Research Foundation, Singapore, and Cyber Security Agency of Singapore.
\end{acks}
\balance
\bibliographystyle{ACM-Reference-Format}
\bibliography{references}

\end{document}